\documentclass[preprint]{iucr}
   \journalcode{J}
\RequirePackage{graphicx}

\usepackage{setspace}
\usepackage{multirow}\usepackage{pbox}
\usepackage{lineno,hyperref}
\modulolinenumbers[5]
\usepackage[dvipsnames]{xcolor}
  \expandafter\let\csname equation*\endcsname\relax
  \expandafter\let\csname endequation*\endcsname\relax

\usepackage{mathrsfs,amsmath}
\usepackage{amssymb}
\usepackage{bm}
\usepackage{dcolumn}
\usepackage{float}
\usepackage{hyperref}
\usepackage{eso-pic}
\usepackage{subcaption}
\usepackage{caption}
\usepackage{capt-of} 
\usepackage{ifthen}
\usepackage{atbegshi}
\usepackage{calc}

\newcommand{\angstrom}{\mbox{\normalfont\AA}}

\usepackage{ tabularx }
\usepackage{ xcolor }
\usepackage{ cancel }
\usepackage{ rotating }
\usepackage{ siunitx }
\usepackage{ xspace }
\usepackage[markup=default]{ changes }
\usepackage{ todonotes }
\usepackage[utf8]{inputenc}
\usepackage{placeins}
\usepackage{soul}

\newcommand{\johanna}{\textcolor{black}} 

\newcommand{\lukas}{\textcolor{black}}

\newcommand{\etal}{\emph{et~al. }}

\newcommand{\corr}{\textcolor{blue}}
\DeclareSIUnit\angstrom{\text {Å}}

\begin{document}

\title{Comparison of time-of-flight with MIEZE neutron spectroscopy of H$_2$O}

\author[a]{L.}{Beddrich}
\cauthor[a]{J. K.}{Jochum}{jjochum@frm2.tum.de}{}
\author[a]{P.}{Bender}
\author[a,b]{L.}{Spitz}
\author[a,c]{A.}{Wendl}
\author[a,d]{C.}{Franz}
\author[e]{S.}{Busch}
\author[b]{F.}{Juryani}
\author[a,c,f,g]{C.}{Pfleiderer}
\author[c,h] {O.}{Soltwedel}

\aff[a]{Heinz Maier-Leibnitz Zentrum (MLZ), Technische Universit\"at M\"unchen, \city{D-85748 Garching}, \country{Germany}}
\aff[b]{Paul Scherrer Institut, \city{CH-5232 Villigen}, \country {Switzerland}}
\aff[c]{Physik Department, Technische Universit\"at M\"unchen, \city{D-85748 Garching}, \country{Germany}}
\aff[d]{J\"ulich Centre for Neutron Science JCNS at MLZ, Forschungszentrum J\"ulich GmbH, \city{D-85748-Garching}, \country{Germany}}
\aff[e]{German Engineering Materials Science Centre (GEMS) at MLZ, Helmholtz-Zentrum Geesthacht GmbH, Garching, Germany }
\aff[f]{Munich Center for Quantum Science and Technology (MCQST), Technische Universit\"at M\"unchen, D-85748 Garching, Germany}
\aff[g]{Zentrum f\"ur QuantumEngineering (ZQE), Technische Universit\"at M\"unchen, D-85748 Garching, Germany}
\aff[h]{Institut f\"ur Physik  Kondensierter Materie, Technische Universit\"at Darmstadt, \city{D-64289 Darmstadt}, \country{Germany}}

\date{\today}

\begin{abstract}
We report a comparison of so called Modulation of IntEnsity with Zero Effort (MIEZE), a neutron spin-echo (NSE) technique, and neutron Time-of-Flight (ToF) spectroscopy, a conventional neutron scattering method. 
Recording the intermediate scattering function $\mathcal{I}(Q, \tau)$ and the scattering function $S(Q, E)$ with MIEZE and ToF, respectively, this comparison involves a Fourier transformation that requires detailed knowledge of the detector efficiency, instrumental resolution, signal background, as well as the range of validity of the spin-echo approximation. 
Discussing these aspects for the properties of pure water as measured on the spectrometers RESEDA and FOCUS under the same experimental conditions, we conclude that an unambiguous understanding of the physical processes dominant in a given system requires information of both, the scattering function as well as the intermediate scattering function. \corr{Furthermore, computational methods like molecular dynamics simulations are essential for understanding these processes and will become increasingly important as we study more complex systems.}
\end{abstract}

\maketitle

\section{Introduction}


Quasielastic neutron scattering (QENS) describes a limit of inelastic neutron scattering, in which energy transfers are small with respect to the energy of the incident neutron.  Three spectroscopic techniques are well established for studies of QENS: backscattering (BS), time-of-flight (ToF), and neutron spin-echo (NSE) spectroscopy.
While these methods provide information on correlations in roughly similar regimes in energy and time as well as momentum and space, the information obtained differs substantially.
For instance, BS and ToF spectroscopy determine the change of the energy of the neutron by means of a crystal analyzer or the time-of-flight of the neutrons, respectively, whereas NSE exploits the precession of the neutron spin in a suitable magnetic field.
Further, BS and ToF provide the dynamic structure factor $S(Q, E)$, while NSE provides the intermediate scattering function $\mathcal{I}(Q, \tau$), where $Q$, $E$, and $\tau$ are the momentum transfers, energy transfers and spin-echo time, respectively.
All methods have their strengths and weaknesses.
Yet conceptual differences in the data collected and the scientific insights gained have been discussed controversially in the scientific community. 

The aim of this paper is to highlight the complementarity of NSE and ToF, and to illustrate key aspects when comparing data. 
As our main message, we argue that an unambiguous understanding of complex physical processes, which are forcibly addressed in cutting-edge research, requires the combination of both $S(Q, E)$ and $\mathcal{I}(Q, \tau$).

Neutron ToF spectroscopy infers the kinetic energy of a neutron from the time-of-flight between two known points in space. 
In indirect geometry ToF instruments, the sample is illuminated by a pulsed white beam and the energy of the scattered beam is determined using a crystal analyzer. 
In comparison, in direct geometry ToF instruments the incident beam is monochromatized either by a crystal monochromator or by a set of at least two choppers.
In both cases the final energy is inferred from the time-of-flight of the distance between the sample and the detector.

ToF spectrometers have the advantage of large detector coverage. 
However, at neutron sources such as reactors, which provide a continuous neutron flux, they suffer from a dramatic reduction in flux due to the chopping of the beam.
In contrast, for pulsed beams provided by spallation sources, ToF instruments appear to be a natural choice. 
ToF instruments are quite versatile with an energy resolution as high as a few $\mu$eV.
Typical instruments cover a large range of momentum transfers, $Q$, that encompass atomic and inter atomic distances below \SI{10}{\angstrom}, but lack resolution when on mesoscopic length scales above \SI{10}{\angstrom} due to their relaxed beam collimation. 

Important examples include LET at ISIS \cite{BEWLEY2011, Nilsen_2017}, or IN5 at the ILL \cite{Ollivier2002}, \cite{Ollivier2011}).
As another advantage, ToF spectrometers may use incident energies of up to 1000\,meV such as MARI \cite{ANDERSEN1996, le2022upgrade}). 
In addition, use of polarized beams as well as polarization analysis are available.

\johanna{Neutron BS utilizes crystals in (nearly) perfect backscattering geometry to analyze the energy of the scattered neutrons. 
The energy of the incoming neutron beam is varied using a Doppler monochromator. 
This configuration offers high energy resolution with a limited dynamic range. 
While instruments at reactor sources offer the highest resolution, with a maximum of 0.3\,$\mu$eV at IN16B at the ILL \cite{gardner_high-resolution_2020, 2010Frick}. 
In comparison, instruments at spallation sources offer a broader dynamical range at a somewhat reduced resolution.
The largest dynamic range of $\pm$ 3.5$\cdot 10^3 \mu eV$ can be reached at the spectrometer IRIS at ISIS \cite{Demmel_2018}.}
Analoguously to ToF, BS spectrometers include a large detector bank, offering simultaneous information on a large range in $Q$ space. 
Instruments with a Si(111) analyzer crystal typically reach up to $Q$ = 1.8\,\AA$^{-1}$ at the highest energy resolutions, while higher $Q$ of up to 3.8\,\AA$^{-1}$ may be reached at a reduced energy resolution, when using a Graphite analyzer \cite{gardner_high-resolution_2020}. 

Taken together, ToF and BS are widely used to study, e.g., molecular reorientation, hydrogen diffusion or liquid dynamics. 
However, ToF and BS cannot resolve the dynamics on mesoscopic length scales such as domain motion in macro-molecules, polymer chain-dynamics, or emergent excitations in quantum magnets.
This shortcoming is due to the combination of intermediate energy resolution and relatively poor resolution of momentum transfers at small $Q$.

Neutron spin-echo techniques are well-known for achieving very high energy resolutions down to below $\SI{1}{\nano\electronvolt}$ \cite{JNSE, IN15}.  Decoupling the energy resolution from the wavelength spread, very high neutron intensity is reached \cite{1972Mezei, 1980Mezei}.
In NSE, comparison of the total phase of the Larmor precession of the neutron spin, acquired in a well-defined magnetic field region before and after the sample, serves to encode energy transfers due to scattering. NSE is especially well established in the investigation of slow relaxation processes on the order of $\sim \SI{1}{\nano\second}$ to $\sim \SI{100}{\nano\second}$. 
Typical scientific problems addressed with NSE are thermal fluctuations of surfactant membranes in microemulsions \cite{Mihailescu2001}, the molecular rheology of polymer melts \cite{Schleger1998}, thermally activated domain motion in proteins \cite{Bu2005}, relaxation phenomena in networks and rubbers, interface fluctuations in complex fluids and polyelectrolytes, transport processes in polymeric electrolytes and gel systems, and the domain dynamics of proteins and enzymes transport process through cell membranes.

To increase the resolution of NSE the field integral seen by the neutron has to be increased.
The associated technical challenges regarding the field homogeneity, have been addressed in different ways.
Most importantly, highly sophisticated correction coils have been developed \cite{Monkenbusch1990}. However, this approach is limited by the energy density stored in these coils and the mechanical forces generated. 
To overcome these limitations of classical NSE instruments, superconducting coils have been developed \cite{2019Pasini, 2009Walter} offering increased field homogeneity and higher magnetic fields. 

Pursuing a different approach to overcome the limitations of classical NSE Golub and Gähler \cite{1987Golub} introduced the resonant neutron spin-echo technique (NRSE), where the solenoids are replaced by a pair of radio-frequency (rf) neutron spin-flippers.
In NRSE, different spin-echo times are reached by tuning the rf flippers to different frequencies. 

While classical NSE and NRSE are well established in studies of soft matter, several shortcomings are known in studies of hard condensed matter.
Perhaps most important, a reduction or total loss of signal occurs in depolarizing samples or sample environments. 
Measurements under such depolarising conditions are cumbersome e.g. as witnessed for so-called ferromagnetic NSE \cite{Mezei2009, Keller2021}.
Further, signal contributions due to incoherent scattering may reduce signal contrast substantially.
In addition, studies in the limit of small momentum transfer $Q$ will suffer from substantial background scattering.

To overcome these limitations of NSE and NRSE, so-called Modulation of IntEnsity with Zero Effort (MIEZE) may be used, representing a variant of NRSE.
Analogous to NRSE \cite{1987Golub, 1992Gaehler}, MIEZE uses rf spin-flippers, instead of large solenoids, to create a precession zone for the neutron spin. 
However, in MIEZE the pair of rf spin-flippers before the sample is operated at different frequencies, leading to an intensity modulation of the signal behind the analyzer, that oscillates with a frequency $f_M = 2\cdot(f_B - f_A)$. 
Placing the analyzer in front of the sample, depolarizing effects by the sample or sample environment no longer affect the measurements.
\corr{Moreover, unlike the polarization in conventional NSE the MIEZE contrast is not reduced by incoherent spin-flip scattering in the sample \cite{1992Gaehler}.}

Since the rf flippers are compact, they allow for the insertion of a field subtraction coil between them \cite{2019Jochum, 2005Haeussler}. 
They permit to extend the dynamic range of MIEZE and NRSE by several orders of magnitude towards shorter echo times, deep into the energy transfers covered by ToF or BS. 
For instance, at RESEDA spin echo times as short as $\tau_\mathrm{min}$ = \SI{0.1}{\femto\second} may be reached at a \corr{wavelength} of \SI{4.5}{\angstrom} \cite{2022Jochum}.
However, this advancement towards shorter spin-echo times, and therefore putatively larger energy transfers, reaches well beyond the spin-echo (SE) approximation, representing the standard framework underlying NSE in the limit of small energy transfers \cite{2019Franz2, 2019Franz3}. 

In this paper we consider the validity of neutron spin-echo techniques in parameter regimes outside the SE approximation. 
Using established numerical methods conceptional challenges implied by the SE approximation remain, such as the energy and wavevector transfer which are not known. 
These limit MIEZE investigations to dispersionless excitations unless additional sample information, such as molecular dynamics (MD) simulations or input from other neutron spectroscopic methods, are available \cite{Keller2021}.


\section{Theoretical framework}

NSE techniques are based on the precession of the neutron spin in magnetic fields as a probe that allows to infer energy transfers during scattering events.
For an introduction we refer to the book by Mezei \cite{mezei2002neutron}.
In the following we focus on the concepts needed to discuss the SE approximation and its implications.
We start by defining the precession angle of a neutron travelling with a velocity $v$ perpendicular to a magnetic field of field strength $B$, and length $L$:
\begin{flalign}
	&\Phi = \Phi_0 + \frac{\gamma B L}{v}, 
\end{flalign}
where $\gamma$ is the neutron's gyromagnetic ratio ($\gamma=\SI{183.25}{\mega\hertz\per\tesla}$ for angles in \corr{radian}).
Without loss of generality, we assume $\Phi_0 = 0$ in the following.
In an NSE setup, a neutron travels across a well defined magnetic field region ($B_1$, $L_1$) before reaching the sample, followed by a trajectory across a second well defined field region after the sample position ($B_2$, $L_2$).
The phase $\Phi_D$ of the neutron at the detector position may be written as: 
\begin{flalign}
	&\Phi_D = \Phi_1 + \Phi_2 = \frac{\gamma B_1 L_1}{v_1} + \frac{\gamma B_2 L_2}{v_2}. \label{eq:phiD}
\end{flalign}
\corr{Choosing} the lengths and field strengths such that: $L_2$ = $L_1$ = $L$, and $B_2$ = $- B_1$ = $B$. Eq. \ref{eq:phiD} becomes 
\begin{flalign}
	&\Phi_D = \gamma B L \cdot \left( \frac{1}{v_1} - \frac{1}{v_2} \right). 
\end{flalign}
Writing $v_2 = v_1 + \Delta v$, where $\Delta v$ is the change in velocity the neutron undergoes when interacting with a sample positioned between the precession fields, one obtains:
\begin{flalign}\label{PHI} 
	&\Phi_D =  \gamma B L \cdot \left( \frac{1}{v_1} - \frac{1}{v_1 + \Delta v} \right).
\end{flalign}
For a purely elastic scattering process, $\Delta v$ = 0, and therefore $\Phi_D$ = 0.
If the neutron exchanges energy with the sample, $\Delta v$ $\neq$ 0, resulting in a phase shift $\Phi_D$. 
The change in neutron energy during such an interaction may be written as: 
\begin{flalign}
	E &= \frac{m}{2}\cdot(v_2^2 - v_1^2)\\\nonumber
	&= \frac{m}{2}\cdot((v_1 + \Delta v)^2 - v_1^2)\\\nonumber
	&= \frac{m}{2}\cdot(2v_1\Delta v + \Delta v^2)\nonumber
	\end{flalign}
The SE approximation assumes $E \ll \frac{1}{2}mv^2$, i.e, the energy transfer, $E$, is much smaller than the kinetic energy, $\frac{1}{2}mv^2$, of the incoming neutrons. 
Thus $\Delta v \ll v_1$, from which it follows:
\begin{flalign}\label{Etrans}
	&E \simeq m v_1 \Delta v 
\end{flalign}
In turn, $\Delta v$ can be written as:
\begin{flalign}
	&\Delta v \simeq \frac{E}{m v_1}
\end{flalign}
Within the spin-echo approximation $\Delta v + v_1 \simeq v_1$ and therefore:
\begin{flalign}\label{SEtime}
	\Phi_D &\simeq  \gamma B L \cdot \frac{\Delta v}{v_1^2}\\\nonumber
	&\simeq \gamma B L \cdot \frac{E}{m v_1^3}\overset{!}{=} \tau \frac{E}{\hbar}
\end{flalign}
Assuming the SE approximation, Eq. \ref{SEtime} hence defines the so-called spin-echo time $\tau$ as a proportionality factor between the neutron phase, $\Phi_D$, at the detector and the energy transfer, $E$.

Using a neutron polarization analyzer and a neutron detector at a certain scattering angle 2$\vartheta$
the number of polarized neutrons are recorded.
This corresponds to the expectation value of $\cos(\Phi_D$) over scattered neutrons ($\langle \cos\Phi_D \rangle$) in 2$\vartheta$. 
Within the spin echo approximation, $\lambda_\mathrm{i}\,\approx\,\lambda_\mathrm{f}$, the momentum transfer $Q$ is well defined via $2\vartheta$ and the probability of a scattering event with energy transfer $E$ is given by $S(Q, E)$. 
This assumption leads to:
\begin{flalign}\label{SQT}
	P_x = \langle \cos\Phi_D \rangle_{E} &\simeq  \frac{\int_{-\infty}^{\infty} S(Q, E) \cos\Phi_D \mathrm{d}E}{\int_{-\infty}^{\infty}S(Q, E)\mathrm{d}E}\\\nonumber
	&\simeq \frac{\int_{-\infty}^{\infty}  S(Q, E) \cos(\frac{E}{\hbar} \tau) \mathrm{d}E}{\int_{-\infty}^{\infty}S(Q, E)\mathrm{d}E}.
\end{flalign}
The \corr{numerator} of Eq. \ref{SQT} is the cosine Fourier transform of $S(Q,E)$, which represents the real part of the time dependent correlation function, also known as the so-called intermediate scattering function $\mathcal{I}(Q,\tau)$.  
The denominator is the static structure factor $S(Q)$. Therefore:
\begin{flalign}
	P_x = \frac{Re(\mathcal{F}(S(Q, E)))}{S(Q)} = \frac{\mathcal{I}(Q,\tau)}{\mathcal{I}(Q, 0)}
\end{flalign}

Within the SE approximation, this leads to the interpretation that the measured polarization $P_x$ is essentially the energy cosine-transform of $S(Q, E)$ normalized by the static structure factor. Alternatively expressed in the time domain, this represents the intermediate scattering function normalized to its value at zero Fourier time.

Since the SE approximation implies that $\lambda_\mathrm{f} \sim \lambda_\mathrm{i}$ the momentum transfer $Q$ is related to the scattering angle via:
\begin{equation}
    Q = \frac{4\pi}{{\lambda}}\mathrm{sin}\left(\frac{2\vartheta}{2}\right). \label{eq:Qel}
\end{equation}
As a consequence, when determining the probability for the scattering event as the pre-factor (see Eq. \ref{SQT2}) approaches $k_\mathrm{f}/k_\mathrm{i}\approx1$, it is sufficient to consider the dynamic structure factor $S(Q, E)$ only instead of the double differential cross section  $\frac{\mathrm{d}^2\sigma}{\mathrm{d}\Omega\mathrm{d}E}$ \cite{zolnierczuk2019efficient}.

\subsection{NSE beyond the SE approximation}

Assuming a typical wavelength of $\lambda = 6$\,\AA\ for the incoming neutrons corresponding to a kinetic energy of $E_{i} = 2.27$\,meV, the SE approximation holds true for quasi-elastic scattering processes with energies in the $\mu$eV-range (see Fig.\,\ref{fig:Fig1}).
Accordingly, NSE is not suited to investigate processes with energy transfers in the meV-range, including inelastic processes. 
$\mathcal{I}(Q,\tau)$ as measured typically is not normalized to the full integral $\int_{-\infty}^\infty {S(Q,E)dE}$ but to a structure factor where the integral is taken over the band-pass of the spectrometer \cite{Richter1998} only.
The lower boundary of this integral is given by $-E_\mathrm{i}$, the energy of the incoming neutron, and the upper boundary is given by the maximum wavelength accepted by the neutron analyzer. 

To go beyond the SE approximation the explicit expression for the phase at the detector $\Phi_D$ must be taken into account (see Eq.\,\ref{PHI}).
In addition, the conservation of energy and momentum must be considered.
Furthermore, the scattering angle $2\vartheta$ must be used instead of $Q$ and the dynamic structure factor $S(Q, E)$ must be replaced by the double differential scattering cross section $\frac{\mathrm{d}^2\sigma}{\mathrm{d}\Omega\mathrm{d}E}$.
Taken together Eq.\,\ref{SQT} then becomes
\begin{flalign}\label{SQT2}
	P_x = \langle \cos\Phi_D \rangle_{E} 
	&=  \frac{\int_{-\infty}^{\infty} \frac{\mathrm{d}^2\sigma}{\mathrm{d}\Omega \mathrm{d}E} \cos\Phi_D \mathrm{d}E}{\int_{-\infty}^{\infty} \frac{\mathrm{d}^2\sigma}{\mathrm{d}\Omega \mathrm{d}E} \mathrm{d}E}\\
	&=\frac{\int_{-\infty}^{\infty}\frac{k_\mathrm{f}}{k_\mathrm{i}}S(2\vartheta, E) \cdot \mathrm{cos}\left( \gamma B L\left( \frac{1}{v_1 + \Delta v} - \frac{1}{v_1} \right)\right)\mathrm{d}E}{\int_{-\infty}^{\infty}\frac{k_\mathrm{f}}{k_\mathrm{i}}S(2\vartheta, E) \mathrm{d}E}
	.
\end{flalign}
\corr{Here, $v_1$ is $\sqrt{\frac{E_{i}}{2m}}$ with the incomming neutron energy $E_{i}$ and $\Delta v$ corresponds to $\sqrt{\frac{\Delta E}{2m}}$, where $\Delta E$ is the energy transfer experienced by the neutron.}

\subsection{The MIEZE method}
\label{ssec:mieze_method}

The data presented below was recorded using the Modulation of IntEnsity with Zero Effort (MIEZE) technique \cite{1992Gaehler}. 
Analoguously to neutron resonant spin-echo spectroscopy, in a MIEZE setup the solenoids, that create the spin precession zones, are replaced by pairs of resonant spin flippers \cite{Golub1987}.
For MIEZE the pair of spin flippers after the sample is removed, and the spin flippers before the sample operated at different frequencies, $f_A$ and $f_B$, being separated by  a distance $L_{AB}$.

Analogously to classical NSE the spin phase at the detector may be defined as \cite{1994Golub, 2002Keller}:
\begin{flalign}\label{EqPhiD}
\Phi_D=2\pi f_M t_D - 2\pi f_M\frac{L_{BD}}{v} + 2\pi f_A \frac{L_{AB}}{v},
\end{flalign}
where $t_D$ represents the time-of-flight at which the neutron arrives at the detector, $v$ is the initial neutron velocity, $f_M=2(f_B-f_A)$ the MIEZE frequency and $L_{BD}$ the distance between the second spin flipper and the detector.
The MIEZE detector is placed such, that the velocity dependent terms cancel out, simplifying the equations \cite{1992Gaehler, 2019Jochum}.

Energy transfers $E$ during the scattering event will induce a delay $\Delta t_D$ in the neutron flight time over the distance $L_{SD}$:
\begin{flalign}\label{EqDt}
\Delta t_D=L_{SD}\left(\frac{1}{v_1}-\frac{1}{\sqrt{v_1^2+2E/m}}\right).
\end{flalign}
This leads to a deviation of the spin phase at the detector:
\begin{flalign}\label{EqPhi}
\Delta\Phi_D=2\pi f_M \Delta t_D,
\end{flalign}
which reduces the contrast.

Within the SE approximation, we can rewrite this equation in exact analogy to classical NSE spectroscopy:
\begin{flalign}\label{EqtM}
\Phi_D\simeq\frac{2\pi f_ML_{SD}\hbar}{mv_1^3}\frac{E}{\hbar}\overset{!}{=}\tau_M\frac{E}{\hbar},
\end{flalign}
where the Fourier time $\tau_M$ serves as a proportionality factor between the phase at the detector and the energy transfer. 

Averaging this effect over all possible energy transfers in perfect analogy to the polarization $P_x$ for classical NSE (see. Eq.\,\ref{SQT2}), results in the general expression for the so-called contrast:
\begin{flalign}\label{EqCgen}
C&=\langle \cos\Phi_D \rangle_{E} = \left<\mathrm{cos}(2\pi f_M\Delta t_D)\right>_{E}\\
&=\frac{\int_{-\infty}^{\infty}\frac{\mathrm{d}^2\sigma}{\mathrm{d}\Omega\mathrm{d}E}\mathrm{cos}(2\pi f_M\Delta t_D)\mathrm{d}E}{\int_{-\infty}^{\infty}\frac{\mathrm{d}^2\sigma}{\mathrm{d}\Omega\mathrm{d}E}\mathrm{d}E}\\
&=\frac{\int_{-\infty}^{\infty} \frac{k_\mathrm{f}}{k_\mathrm{i}}S(2\vartheta, E)\mathrm{cos}\left(2\pi f_ML_{SD}\left(\frac{1}{v_1}-\frac{1}{\sqrt{v_1^2+2E/m}}\right)\right)\mathrm{d}E}{\int_{-\infty}^{\infty} \frac{k_\mathrm{f}}{k_\mathrm{i}}S(2\vartheta, E)\mathrm{d}E}.
\end{flalign}


\subsection{Computation of the MIEZE contrast}
\label{ssec:MIEZEcomputation}

\begin{figure*}
    \centering
    \includegraphics[width=\textwidth]{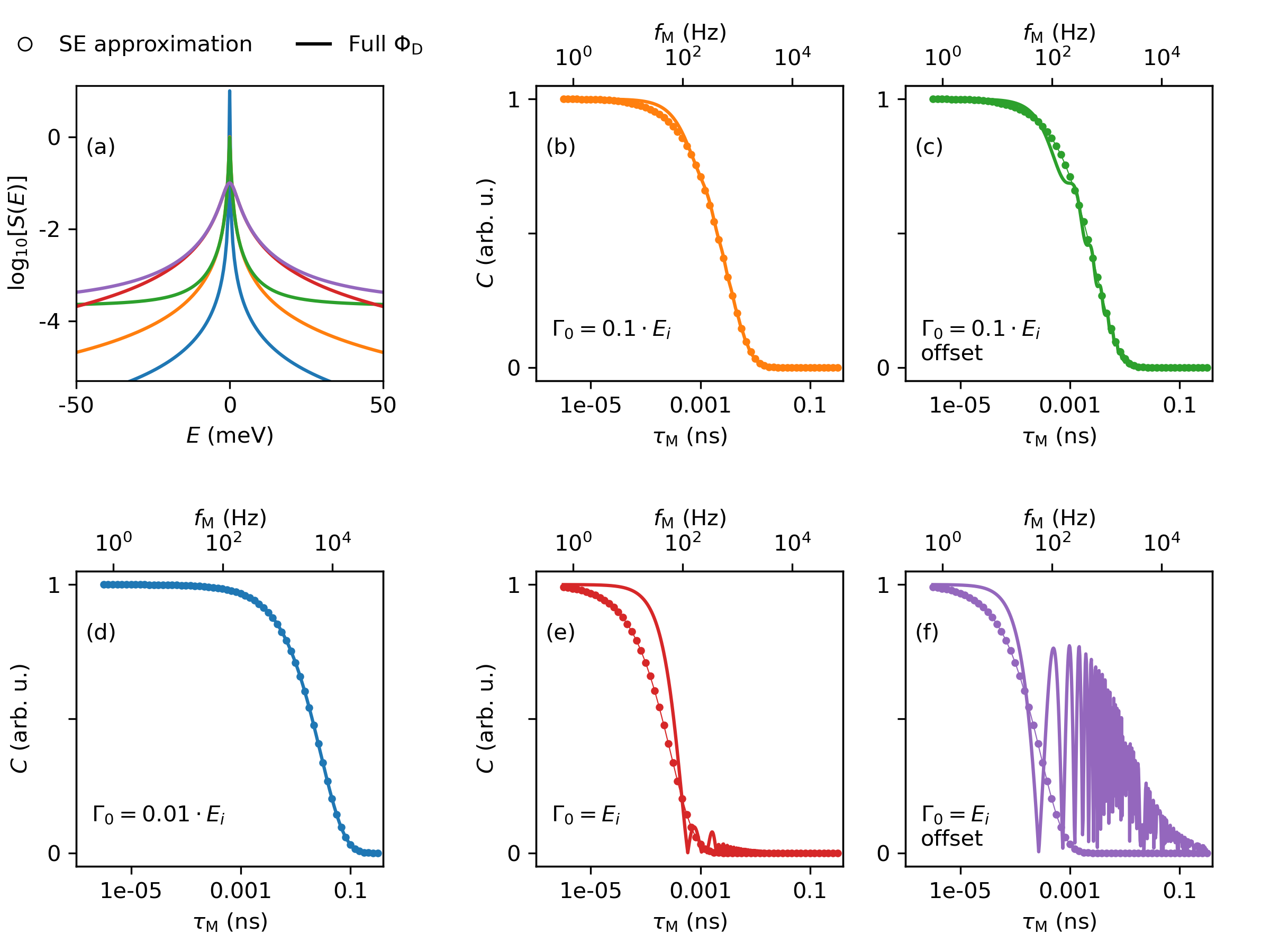}
    \caption{\label{fig:Fig1} Visualization of the difference between spin-echo approximation and the explicit phase calculation. 
    Panel (a) shows five dynamic structure factors of quasielastic processes whose transformation into the time domain is shown in graphs (b)-(f) as indicated by their color.  
    In (b)-(f) the transformation calculated in SE approximation is shown by circles, while the solid lines depict the result of Eq. \ref{EqCgen}. 
    The wavelength of the incoming neutron beam was assumed to be monochromatic and $\lambda = \SI{6.0}{\angstrom}$, \corr{which corresponds to $E_\mathrm{i} = $ \SI{2.27}{\milli\electronvolt}.}}
\end{figure*}

To elucidate the physical meaning of the contrast $C$, we examine a few illustrative cases. To evaluate the impact of the SE approximation, we computed the expected MIEZE contrasts for different dynamic structure factors $S(Q, E)$. 
We assumed a monochromatic neutron beam with wavelength $\lambda=6$\AA, corresponding to a kinetic energy of \corr{$E_\mathrm{i} =$} \SI{2.27}{\milli\electronvolt}. 
Fig.\,\ref{fig:Fig1}(a) shows five Lorentzian distributions $S(E)\propto\Gamma_0/(\Gamma_0^2+E^2) + BGRD$, \corr{representing quasielastic scattering,} where $\Gamma_0$ is the distribution’s linewidth \corr{(HWHM)}, and $BGRD$ denotes a potential constant background. Within the SE approximation, the MIEZE contrast is computed as a cosine Fourier transform of $S(E)$.

Applying Eq.\,\ref{SQT} to the Lorentzian distributions in Fig.\,\ref{fig:Fig1}(a) produces the curves shown by circles \corr{in corresponding colors} in Figs.\,\ref{fig:Fig1}(b)-(f). 
These curves exhibit an exponential decay with $C\propto \exp(-\Gamma_0\tau_M / \hbar)$. \corr{Using Eq.\,\ref{EqCgen} — applicable independently of the measured excitation energy — to calculate the contrast from the curves in Fig.\,\ref{fig:Fig1}(a), yields the continuous lines, also color-matched.
Here, $E_\mathrm{i}$ represents the incoming neutron energy. 
For a linewidth of 1\,\% of $E_\mathrm{i}$, the SE approximation and the explicit calculation align perfectly (Fig.\,\ref{fig:Fig1}(d)). 
Increasing $\Gamma_0$ to 10\,\% of $E_\mathrm{i}$ introduces minor deviations (Fig.\,\ref{fig:Fig1}(b)), while $\Gamma_0 = E_\mathrm{i}$ produces significant discrepancies. 
Furthermore, adding a constant background (Figs.\,\ref{fig:Fig1}(c) \& (f)) introduces an oscillatory modulation atop the exponential decay.
These oscillations, which arise for large $\Gamma_0$ or increased background, are due to the nature of the finite integration window of the Fourier transform \cite{Wuttke1995}}. 
\corr{In first approximation the finite energy window, bounded below by maximum neutron energy loss and above by the highest detectable neutron energy, can be thought of as a rectangular function. 
Therefore, we are effectively performing a Fourier transform of $S(2\vartheta, E)$ convoluted with a rectangular function. 
As the Fourier transform of a rectangular function is the $sinc$ function, this results in a contrast $C$, modulated with a $sinc$ function, leading to damped oscillations along the positive x - axis, with an amplitude that decreases as $\frac{1}{n}$. 
Here, $n$ is defined via the Taylor series $\mathrm{sinc(x)} = \frac{\sin(x)}{x} = \sum_{n=0}^\infty\frac{(-1)^n x^{2n}}{(2n +1)!}.$} 
These oscillations are absent in the SE approximation (see Fig.\,\ref{fig:Fig1}(c)). 
Consequently, oscillations indicative of purely inelastic scattering in the SE approximation \corr{may also be present when measuring a quasielastic scatterer.} 
These two contributions can not easily be disentangled, \corr{noting} that the parameters used for this illustration in Figs.\,\ref{fig:Fig1}(e) \& (f) are quite extreme.\\

\begin{figure*}[h]
    \centering
    \includegraphics[width=\textwidth]{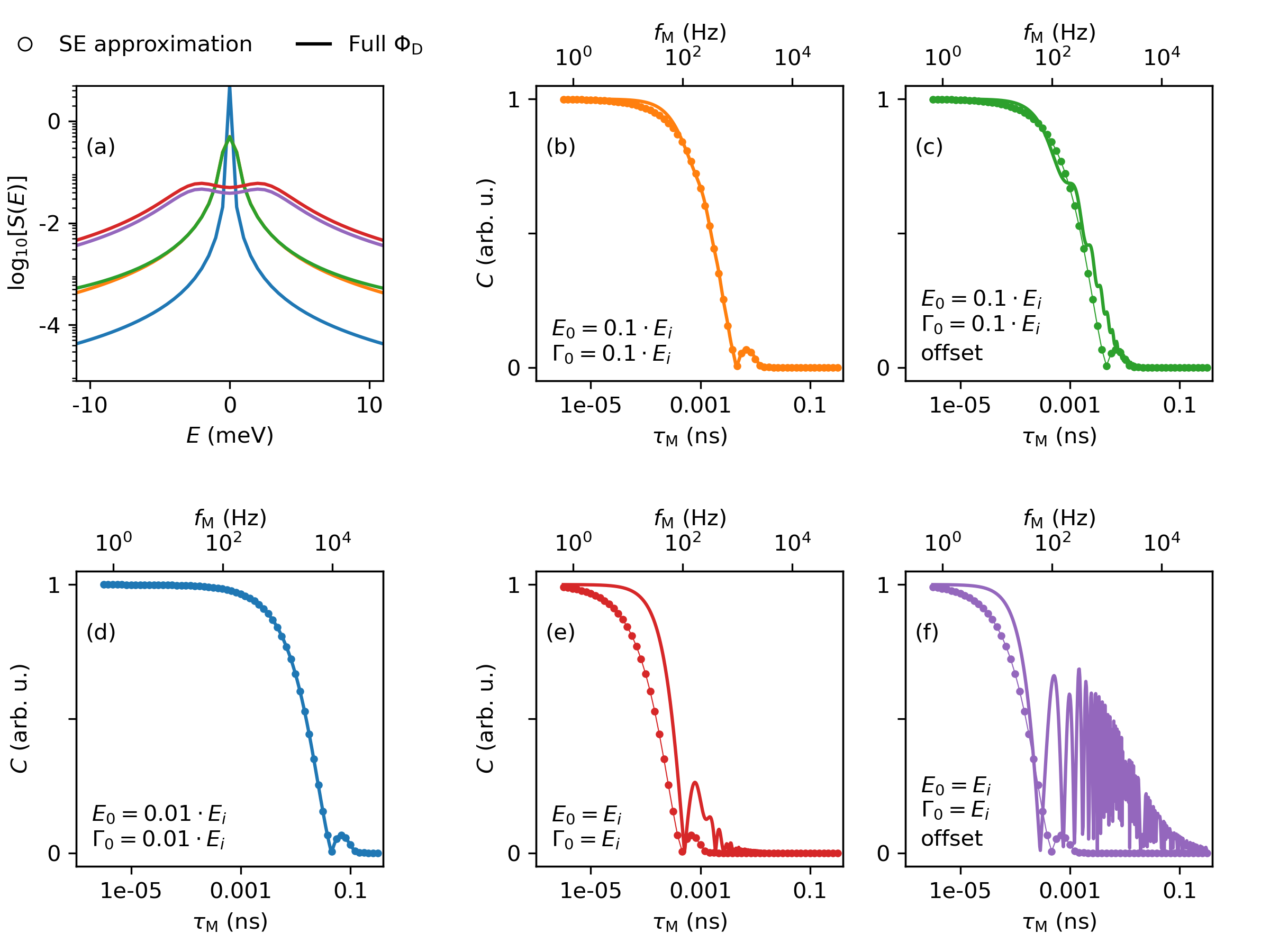}
    \caption{\label{fig:Fig1ins} Visualization of the difference between spin-echo approximation and the explicit phase calculation. 
    Panel (a) shows five dynamic structure factors of inelastic processes whose transformation into the time domain is shown in graphs (b)-(f) as indicated by their color.  
    In (b)-(f) the transformation calculated in SE approximation is shown by circles, while the solid lines depict the result of Eq. \ref{EqCgen}. 
    The wavelength of the incoming neutron beam was assumed to be monochromatic and $\lambda = \SI{6.0}{\angstrom}$, \corr{which corresponds to $E_\mathrm{i} = $ \SI{2.27}{\milli\electronvolt}.}}
\end{figure*}

\corr{Fig.\,\ref{fig:Fig1ins} visualizes, in the same layout of Fig.\,\ref{fig:Fig1}, true inelastic scattering.
Fig.\,\ref{fig:Fig1ins}(a) shows five Lorentzian distributions for finite excitation energies $E_0$: $S( E)\propto(\Gamma_0/(\Gamma_0^2+(E + E_0)^2) +\Gamma_0/(\Gamma_0^2+(E - E_0)^2) + BGRD$, where $\Gamma_0$ and $BGRD$ are defined as above.
Eq.\,\ref{SQT} applied to the Lorentzian distributions in Fig.\,\ref{fig:Fig1ins}(a) leads to curves that comprise an  exponential decay with $C\propto \mathrm{exp}(-\Gamma_0\tau_M / \hbar)$ modulated by a $\mathrm{cos}(E_0 \tau_M/\hbar)$-term.
Similarly to the quasielastic case, the SE approximation and the explicit calculation (Eq.\,\ref{EqCgen}) agree well if $E_0$ and $\Gamma_0$ are smaller than 10\,\% of $E_\mathrm{i}$. 
For increasing $\Gamma_0$ and $E_0$ the deviations increase as well, leading first to a mismatch in curvature of the decay and later on to additional oscillations for Eq.\,\ref{EqCgen}. 
Analogously to the quasielastic case we can see that an increased background leads to additional oscillations. }

\begin{figure*}[h]
    \centering
\includegraphics[width=\textwidth]{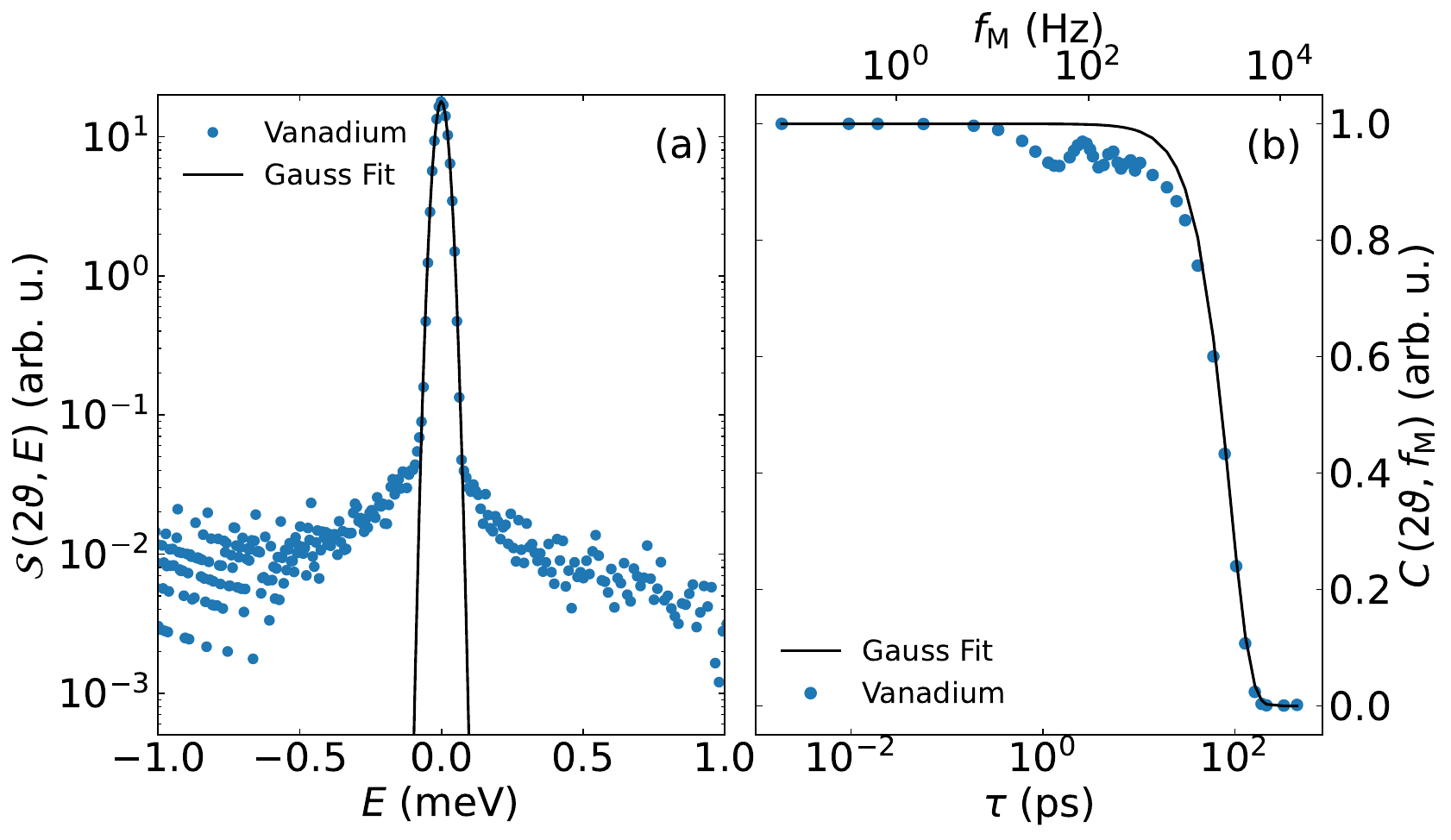}
    \caption{\label{fig:Vanadium} \corr{Vanadium data recorded at TOFTOF. (a) Vanadium data in energy space is shown together with a Gaussian fit to the data (b) The datasets shown in (a) after the MIEZE transform (Eq.\,\ref{EqCgen}) is applied.}} 
\end{figure*}

\corr{The features discussed here, though presented in an exemplary manner, are indeed observed in real data.
They must be taken into account when analyzing MIEZE and classical NSE data, as well as when applying a Fourier transform to ToF data to separate contributions to the resolution function \cite{ZORN2012}.
In ToF measurements, the resolution function ($ R(Q, E)$), which comprises contributions from incident energy spread, initial pulse width, flight path length uncertainties, sample geometry, detector depth, etc., enters the experimentally measured $S(Q, E)_{exp}$ as a convolution: $S(Q, E)_{exp} = R(Q, E) \ast S(Q, E)$ \cite{ZORN2012}.
Using the convolution theorem, the Fourier transform of this expression simplifies to: $S(Q, t)_{exp} = R(Q, t) \cdot S(Q, t)$, making it in principle possible to deconvolute the measured data from resolution effects.
This procedure follows the same rules as those used for NSE and MIEZE: energy dependent quantities like detector efficiency \cite{2016Koehli}, polarization analysis, transmission coefficients, in general, all contributions to the instrumental resolution function need to be considered.
As we have seen in the examples in Figs.\,\ref{fig:Fig1}\,\&\,\ref{fig:Fig1ins}, a constant background can have a significant effect on the Fourier transform of $S(2\vartheta, E)$, and can lead to a mismatch when comparing ToF data to NSE or MIEZE data (see further Sec.\,\ref{sec:data_analysis_discussion}).} 
\corr{Considering the Vanadium data, shown in Fig.\,\ref{fig:Vanadium}, recorded at TOFTOF we can see that while a Gaussian fit to this data leads to a clean curve when the MIEZE transform (Eq.\,\ref{EqCgen}) is applied, the data itself suffers from osciallations that originate in the instrumental background, that stretches over a wide range in energy, up to the cutoff for our integration window.}

\section{Experimental Methods and Data Reduction}
\label{sec:experimental_methods}

The intermediate scattering function of liquid Milli-Q\texttrademark ultrapure water was measured using the MIEZE option at the resonant spin-echo spectrometer RESEDA \cite{reseda, 2019Franz, 2019Franz2} at the MLZ, \corr{under a scattering of angle $2\vartheta$ = 22$^\circ$}.
For the measurements, the water was kept in a Hellma macro cell with an optical path length of \SI{1}{\milli\meter}.
\corr{The cell thickness was chosen to optimize the scattering signal. Previous measurements, using a Hellma macro cell with an optical path length of 0.2\,mm have suffered from low statistics. However, in terms of evaluated translational diffusion times the data extracted from the measurements using the different cells agreed well with each other and with literature values. Since the very same sample cells  were utilized for MIEZE and ToF-spectroscopy a direct comparison between both techniques is practicable even in the presence of multiple scattering.}
The cell was kept at a temperature of \SI{300}{\kelvin} using a closed cycle cryostat in closed loop temperature control. 
The resolution curve was recorded using carbon powder, representing a purely elastic scatterer, inside an identical Hellma macro cell.
The complementary ToF measurement was performed at the instrument FOCUS at the PSI \cite{Janssen1997, Mesot1996} \corr{using the same sample}.
Here, we measured the energy spectra $S(2\vartheta, E)$ at $T=300$\,K for a large angular range.
\corr{Preliminary measurements using an a cylindrical aluminum container with outer diameter d$_1$ = 22.5\,mm and inner diameter d$_2$ = 22.2\,mm, and therefore an effective sample thickness of 0.15\,mm were performed at the spectrometer TOFTOF at the MLZ \cite{UNRUH2007}. The influence of the sample geometry on the comparability of MIEZE and ToF data are shown below.}


The neutron scattering data were reduced using standard procedures. In case of the MIEZE data, the reduction was done using the software package MIEZEPY \cite{MIEZEPY}. 
\corr{This procedure includes the normalization of the sample contrast to the contrast of a purely elastic scatterer to take into account all instrument specific resolution effects, such as the different wavelengths and wavelength bands used, potential polarization losses in the primary spectrometer, scattering off sample environment and intrinsic background scattering off instrument components.}
The process yields the contrast $C(2\vartheta, f_\mathrm{M})$ at a fixed scattering angle $2\vartheta$.

The ToF data was normalized to the incoming flux to account for small fluctuations. 
To account for scattering off the cuvette, an empty cell was measured and the data subsequently subtracted. 
The detector was calibrated using a vanadium standard, taking into account a correction for the anisotropy of the Debye-Waller-factor of vanadium. 
The vanadium measurement was also used to determine the resolution of the elastic line. 
Subsequent to these corrections the energy transfer was calculated from the measured neutron time-of-flight. 
\corr{In the case of TOFTOF}, the data was binned in $E$, leading to a small energy offset of \SI{-4.53 \pm 0.22}{\micro\electronvolt}. 
The elastic line determined in the measurements of water was corrected by this offset.

Knowing the energy transfer and the scattering angle $2\vartheta$ the momentum transfer $Q$ may be calculated. 
However, this step was omitted since a comparison with the data recorded at RESEDA was only possible for constant scattering angle $2\vartheta$. 
As a final step, the measured neutron intensity was corrected for the energy dependent detector efficiency \cite{UNRUH2007} as well as the $k_\mathrm{f} / k_\mathrm{i}$ factor in the double differential cross section.

\corr{For the data acquired at FOCUS, the reduction was performed using the \textit{Data Analysis and Visualization Environment} (DAVE \cite{Azuah2009}), while TOFTOF data was treated with \textit{Mantid}\cite{ARNOLD2014}, applying the same processing and correction steps as outlined before.}


\section{Data analysis}
\label{sec:data_analysis_discussion}

Two separate approaches were pursued to compare the data measured at the three different spectrometers.
(i) We compared the acquired energy spectra via a direct transformation into the time / frequency domain with the RESEDA data.
(ii) Due to the stark deviations in the comparison between TOFTOF and RESEDA also in the range of 'medium high' energy transfer, we fitted a heuristic $S(2\vartheta, E)$ model to both data sets taking into account models for the instrument resolution, bandwidth and energy dependent detector efficiencies.
Ultimately, the discrepancy found in approach (i) was reduced, when re-measuring the same Hellma macro cell at FOCUS, hinting on the profound influence of sample geometry on the direct comparability of the MIEZE and TOF technique.

\begin{figure*}[htp]
\begin{center}
     \includegraphics[width=0.8\textwidth]{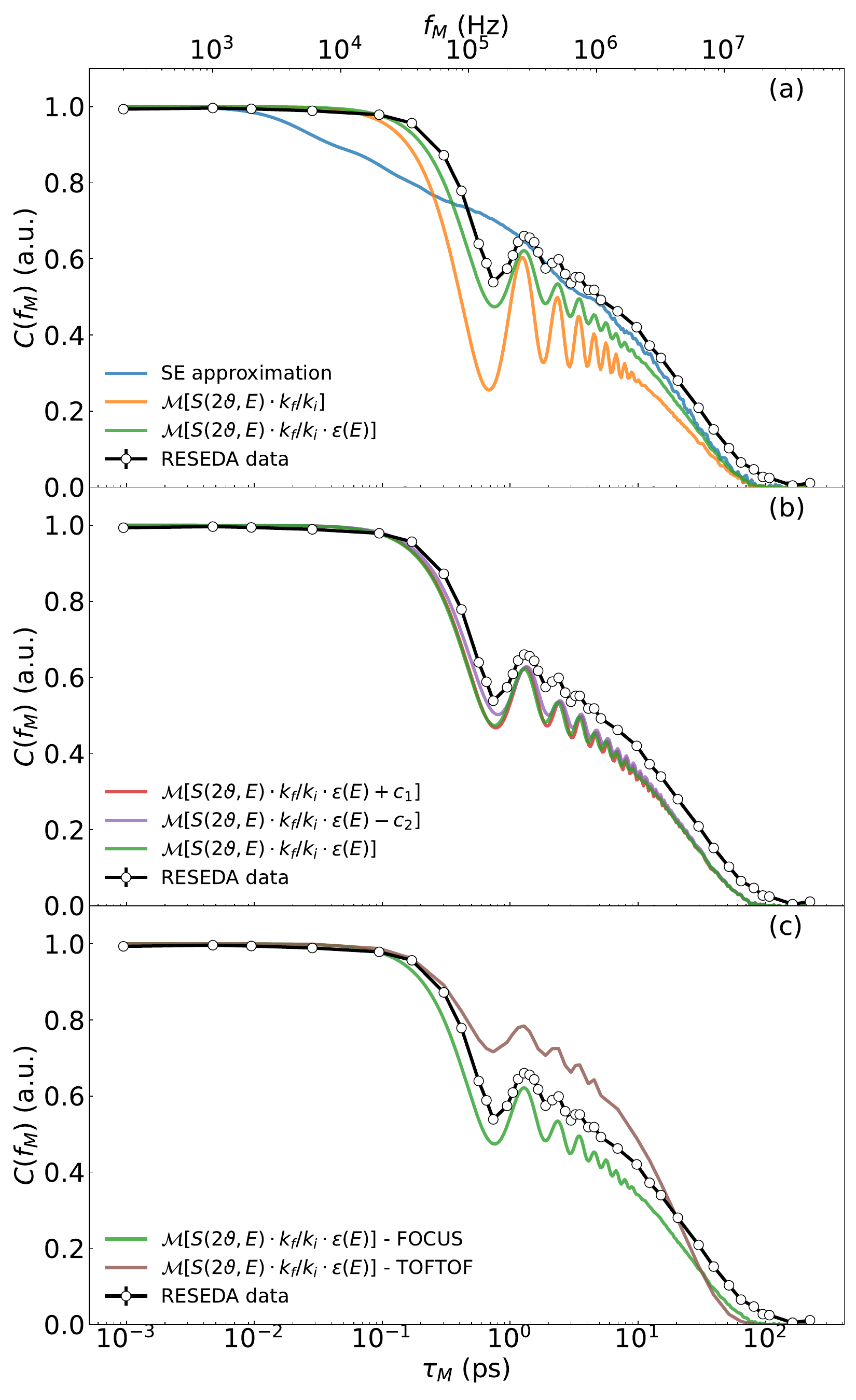} 
     \caption{Direct transformation of TOFTOF and FOCUS spectra $S(2\vartheta\!=\!\SI{22}{\degree}, E)$ into the time domain. In (a) we show the transformed time-of-flight spectrum of FOCUS with increasing number of corrections for direct comparison with the RESEDA data (circles). The simplest case uses the spin-echo approximation (Eq. \ref{SQT}) and results in the blue curve. The orange curve is obtained by transforming the energy spectrum}
        \label{fig:direct_transformation}
\end{center}
\end{figure*}

As motivated in Sec. \ref{ssec:mieze_method} and \ref{ssec:MIEZEcomputation} it proves to be essential to go beyond the SE approximation to transform a ToF energy spectrum. 
The results of applying Eqs. \ref{SQT}, and \ref{EqCgen} to the data measured on TOFTOF and FOCUS in comparison to the data measured on RESEDA (circles) can be seen in Fig. \ref{fig:direct_transformation}.
The errorbars representing the statistical uncertainty in the contrast value are smaller then the marker itself.
In Fig. \ref{fig:direct_transformation} (a) an increasing number of corrections to the simple SE approximation were taken into account in the transformation of the FOCUS data.
Emphasizing the inadequacy of the SE approximation, the blue line has been calculated using the standard framework expressed by Eq. \ref{SQT}, which is unable to describe the MIEZE data.
Adding the full expression of the MIEZE phase change $\Delta \Phi_D$ as well as the correction factor $k_\mathrm{f} / k_\mathrm{i}$, introducing a natural cut-off at $k_\mathrm{f} = 0$, yields the orange curve that reproduces the characteristic oscillations in terms of their frequency but not with the correct amplitude.
Factoring in the energy dependent detection efficiency of the CASCADE detector at RESEDA \cite{2016Koehli}, designated as $\varepsilon(E)$, achieves a reduction in the amplitude of the oscillations around \SIrange{100}{1000}{\hertz} to the exact level seen in the MIEZE data (green curve). This transformation including the full expression of the MIEZE phase, the factor $k_\mathrm{f} / k_\mathrm{i}$, as well as the CASCADE detector efficiency will be referred to as "MIEZE transformation" or simply "transformation" in the following.

Despite these improvements, it is unfortunately not possible to reproduce the exact numerical values of the MIEZE data, neither in the mid frequency range nor in the high frequency range.
A constant offset between the RESEDA data and the transformed FOCUS data is visible.
Both discrepancies ought to have different origins, since the oscillations were affected by $\varepsilon(E)$, whereas the faster decay at high $f_\mathrm{M}$ was unaffected. 
Most notably, the faster decay is present in the SE approximation as well.

The main source for the remaining mismatch in the high frequency range, $f_\mathrm{M} > \SI{1000}{\hertz}$, is expected to be the fact that the time-of-flight data is the convolution of the sample's response with the instruments resolution function.
Mathematically, it is not possible to disentangle the ToF data from the instrumental resolution and acquiring a unique solution.
An inverse Fourier transformation based on Bayesian analysis, employing various regularization techniques have successfully been used in, e.g., the analysis of small-angle neutron scattering data \cite{bender2017structural}, have failed thus far.
Mostly due to the low density of data points and the reduced sensitivity of the MIEZE technique to high energy inelastic scattering, numerical artifacts lead to nonphysical solutions insurmountable by regularization.
In that regard, we cannot capitalize from the advantage of the MIEZE technique, where the instrumental resolution in the time domain can easily be disentangled from the data by normalizing it to the data of a purely elastic scatterer.

However, an approach of simultaneously fitting a data set consisting of MIEZE and ToF can still be employed.
Starting from a $S(2\vartheta, E)$ model, the ToF data can be fitted by taking an analytical model of the instrument resolution function \cite{2007Gaspar} into account, while the MIEZE data is compared to the MIEZE transform of the same model.
For example, the parameters of a heuristic model comprising a sum of multiple Lorentzian functions have been optimized against the TOFTOF and RESEDA data in a combined $\chi^2$ fit.
Due to the sensitivity of the MIEZE method to the flanks of $S(2\vartheta, E)$ appropriate model functions are required to extract the correct, physical parameters from the data. 
In the context of critical phenomena, correlation functions are known to assume more complex shapes \cite{boni_comparison_1993, beddrich2023consequences} and the assumption of simple Lorentzian distributions in energy space breaks down.
Since this has not been the focus of the results we want to present, the fitting procedure and findings are described in appendix \ref{app:combined_fit}.

\clearpage
\captionof*{figure}{ with Eq. \ref{EqCgen}, which reproduces quite accurately the frequency of the oscillations visible in the RESEDA data. In green, we have also considered the energy dependent detection efficiency of the CASCADE detector, which now accounts for the correct amplitude of the oscillations. Comparing the latter with the MIEZE data, at high frequencies the transformed ToF data decays faster and at medium frequencies the contrast is about \num{0.05} lower than expected. Plot (b) visualizes the effect of adding (red line) or subtracting (purple line) a constant background, leading to a downward and upward shift compared to the original curve, respectively. In (c) we show the comparison between the RESEDA (circles), FOCUS (green line) and TOFTOF (brown line).
The differences between FOCUS and TOFTOF, which ought to account for the discrepancies, are the instruments themselves, the usage of their respective standard data reduction software and the shape of the sample container.}

For the mid frequency range, however, the one factor contributing to the difference between the RESEDA and the transformed ToF data appears to be due to the background of the energy spectrum and its treatment.
This can be seen in panel \ref{fig:direct_transformation} (b), where we compare the MIEZE transformation of the FOCUS data (green line) with the same data set that has been modified by adding an additional constant background $c_1 = 0.058$ or by subtracting $c_2 = 0.035$.
This seems minuscule compared to the peak value of the measured intensity at FOCUS $S(2\vartheta, E)_\mathrm{max} = \SI{1593}{(\mathrm{arb. u.})}$.
However, the resulting contrast curve has been noticeably shifted upwards (red dotted line) and downwards (red solid line) in the mid to low frequency range, respectively, while general structure of the oscillations have been preserved.
When adding a constant background, the contrast shifts down, because in relative terms, more integrated intensity is added to the high energy part of the spectrum.
Integrated over the entire energy range of $S(2\vartheta, E)$, the contribution of the constants $c_1$ and $c_2$ accounts for about 2.2 \% and 1.3\% of the spectrum, respectively.
This shows that for a successful comparison with MIEZE data, a detailed knowledge of the instrument background, especially at high energies, is essential.
Spurious, inelastic scattering at such high energy transfers cannot be ruled out, but seems to be unrealistic.
The dark count rate of the ToF detectors could prove to be crucial. 
Dark counts are equally distributed over the time bins and appear as a constant background along the energy band resolved. 
These finite values at the edges of the Fourier-transform window alter the amplitude of the oscillations in the time domain drastically, as shown in Fig.\,\ref{fig:direct_transformation}\,b) as well as Figs.\,\ref{fig:Fig1} and \ref{fig:Fig1ins}.
In this context, we suspect that incomplete knowledge about the efficiency of the detectors at high neutron energies may be complicating the issue. 

\corr{Furthermore, sample geometry has been identified as a major pitfall for comparing ToF and MIEZE data directly. Fig. \ref{fig:direct_transformation} (c) compares the MIEZE transformation of the FOCUS and TOFTOF data, which have been acquired using two different geometries of the sample container.
As described in section \ref{sec:experimental_methods}, the sample containers used in the TOFTOF and FOCUS measurement had a cylindrical and rectangular, plate-like shape, respectively.
The difference between the two cases is striking as the solid green curve again shows the transformed FOCUS data, while the brown line represents the transformed TOFTOF data.}
\corr{A detailed analysis of the influence of different sample shapes on ToF data has been performed by Zorn \etal \cite{ZORN2012}, and would go beyond the scope of this manuscript.}

\section{Conclusions and Outlook}

In conclusion, we have introduced a transformation, which enables us to directly compare any excitation spectrum as measured in a ToF instrument or calculated from solid state theory with its counterpart observed by a MIEZE spectrometer in the time domain.
We take several factors into account that become important when considering large energy transfer with respect to the kinetic energy of the neutron.

We successfully applied this framework to data acquired on an ultra-pure water sample at room temperature using the ToF spectrometers FOCUS and TOFTOF to compare the results with the data measured at the MIEZE instrument RESEDA.
In the case of identical sample conditions, the agreement between the transformed FOCUS data and RESEDA data is good.
However, we identified several pitfalls that significantly affect the transformation, complicate direct comparisons, and necessitate careful consideration.
Somewhat unwillingly, the TOFTOF data provide an excellent illustration of this point, due to the differing geometry of the sample container.

In addition to the geometry of the sample, effects such as energy resolution of the ToF spectrometer, character of the background and control over the latter during the data reduction process influences the transformation.
Dark counts in the detector electronics, spurious scattering from e.g. sample environment components or air and cross talk due to coarse collimation are certainly difficult to characterize often leaving a background that cannot be neglected.
As a final significant influence, accurate knowledge of the neutron detection efficiency, spin-flip ratio of the analyzer, and its dependence on the neutron energy is required when analyzing data beyond the SE approximation.

Having these insights in mind our transformation can be employed in various approaches to analyzing and planning experiments.
Besides directly transforming recorded energy spectra into the time domain, a combined fitting procedure for ToF and MIEZE data can be employed. For example, the high resolution of MIEZE can constrain slow dynamics, while ToF provides context over a broader range of energy and momentum transfer.
We have demonstrated this in appendix \ref{app:combined_fit} by fitting a heuristic model comprised of a sum of several Lorentzian distributions.
Here, we can account for resolution effects of time-of-flight instruments in the form of an analytical resolution model.

\begin{figure*}[h]
    \centering
    \includegraphics[width=0.8\textwidth]{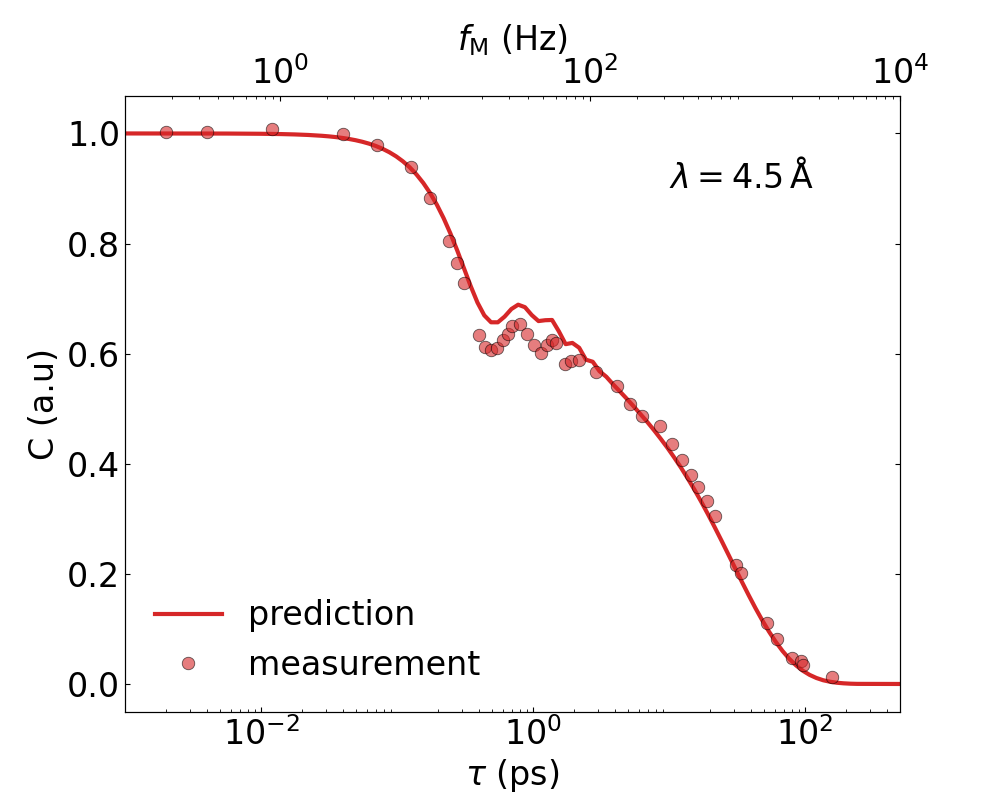}
    \caption{\label{fig:prediction} Comparison of the predicted contrast curve calculated from the optimized 3Q1I model to the corresponding RESEDA data at $\lambda_\mathrm{i} = \SI{4.5}{\angstrom}$. The data was collected at the appropriate angle $2\vartheta$ such that momentum transfer $Q$ is the same in the SE approximation. The predicted curves accounts for the increased integration area and the adjustments in the MIEZE phase calculation.} 
\end{figure*}

Taking this model as an example, we want to emphasize the opportunity of identifying unequivocal features to distinguish competing models and determine suitable measurement configurations.
In the context of water, this could be MD simulations with differing force field parameters.
Fig.\,\ref{fig:prediction} presents the result of transforming the model from Fig.\,\ref{fig:3QENS1INS}, which was fitted to the data acquired at $\lambda_\mathrm{i} = \SI{6.0}{\angstrom}$ while accounting for the different $\lambda_\mathrm{i} = \SI{4.5}{\angstrom}$ setting.
The prediction (red line) without any adjustment of model parameters is in good agreement with the measured data, except for the precise vertical position of the oscillations.

\begin{figure*}[h]
    \centering
    \includegraphics[width=0.8\textwidth]{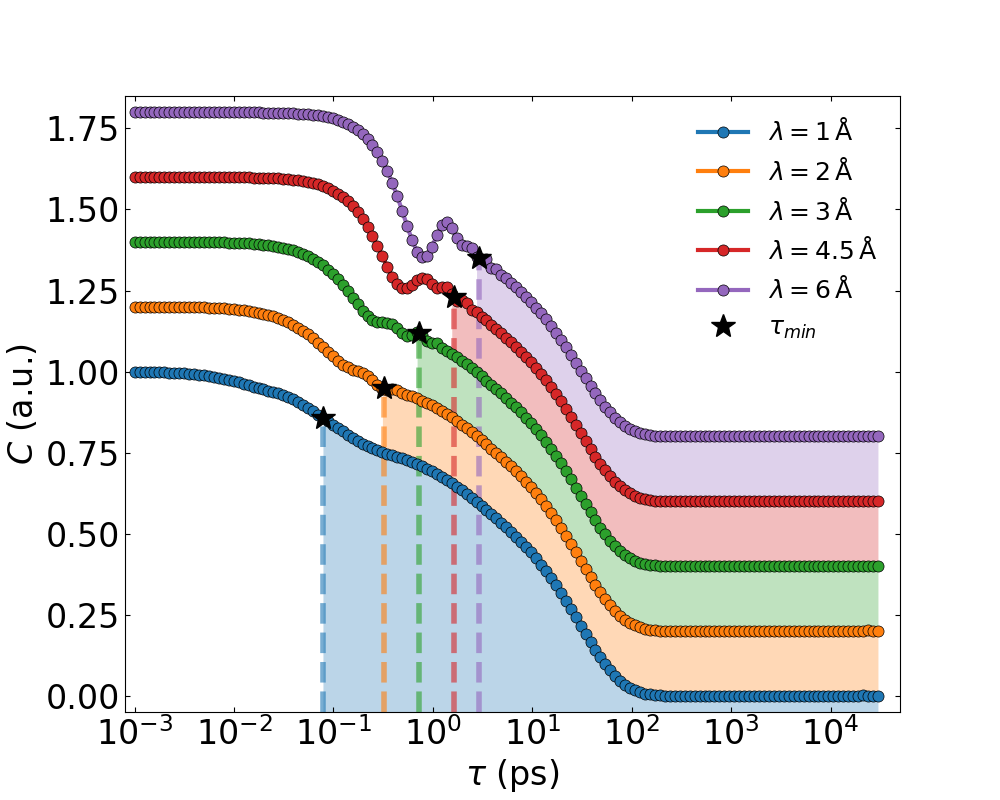}
    \caption{\label{fig:outlook} Prediction for measurements at RESEDA for different initial wavelengths $\lambda_\mathrm{i}$. The purple curve corresponds to the results of fitting the TOFTOF data with model 3Q1I. This result was then used to predict MIEZE data for different wavelengths, similar to Fig. \ref{fig:prediction}. The star labeled $\tau_{min}$ indicates the smallest Fourier time, below which the spin-echo approximation becomes invalid, at the corresponding wavelength $\lambda_\mathrm{i}$.} 
\end{figure*}

The oscillations, which are in this case a consequence of the MIEZE method, are a source of instability in the fitting process and possibly lead to increased uncertainty in model parameters.
However, as demonstrated in section \ref{ssec:MIEZEcomputation}, a higher kinetic energy of incoming neutrons suppresses these unwanted features, increasing reliability of data analysis.
Fig. \ref{fig:outlook} visualizes this effect.
The different curves correspond to a transformation of the same model function as before with decreasing wavelength $\lambda_\mathrm{i}$. The star marks the MIEZE time $\tau_\mathrm{min} = \hbar / (0.1 \cdot E_\mathrm{i})$ below which the SE approximation is no longer valid.
This capability would become available when implementing the proposed instrument upgrade TIGER \cite{2022Jochum} at RESEDA.

\clearpage


\section{Acknowledgments}

We gratefully acknowledge discussion and support from Georg Ehlers, Thomas Keller, Bela Farago, Olaf Holderer, Marcell Wolf, Neslihan Aslan, Dominik Schwaiger, Doaa Ali, and Wiebke Lohstroh. 
We acknowledge also financial support through the BMBF projects ‘Longitudinale Resonante Neutronen Spin-Echo Spektroskopie mit Extremer Energie-Aufl\"{o}sung’ (F\"{o}rderkennzeichen 05K16W06) and 'Resonante Longitudinale MIASANS Spin-Echo Spektroskopie an RESEDA' (F\"{o}rderkennzeichen 05K19W05) and the European Research Council (ERC) through ERC Advanced Grant No. 788031 ‘Extreme Quantum Matter in Solids’ (ExQuiSid).


\appendix
\section{Combined fit}
\label{app:combined_fit}
A common approach to assess neutron data is a standard \textit{forward analysis}, meaning that an idealized model function is assumed and compared with the experimental data. 
Here, a heuristic model is given by a sum of \textit{Lorentzian distributions}
\begin{equation}
    S(2\vartheta, E) = \sum_i \frac{A_i}{\pi} \frac{\Gamma_i}{(E - E_i)^2 + \Gamma_i^2}~,
\end{equation}
where $2\Gamma_i$ is the linewidth (full width at half maximum), $E_i$ is either $0$ or a finite energy transfer, and $A_i$ is the contribution of the excitation to the total dynamical structure factor, normalized to unity ($\sum_i A_i = 1$).

\corr{We have used this analysis scheme to compare the two data sets from TOFTOF and RESEDA that served as the starting point for this study, because the first two approaches of reconciling the data failed.
Initially, we tried to perform an inverse Fourier transformation based on Bayesian analysis, which could not be realized following the aforementioned procedure of Bender \etal \cite{bender2017structural}.
Similarly, the direct transformation of the TOFTOF spectrum did not yield promising results, which was later understood to originate from the differing geometries of the sample container.
Prior to this realization and a subsequent measurement at FOCUS with the correct sample shape, three variations of the heuristic Lorentzian model, consisting of two quasielastic + one inelastic peak (2Q1I), three quasielastic peaks (3Q) and three quasielastic + one inelastic peaks (3Q1I) have been fitted to the combined data set.
Once the FOCUS data was obtained and the direct transformation from the energy to the time domain was successfully performed, the combined fit analysis offered limited additional value in terms of numerical outcomes. However, we find it instructive to include the procedure and sharing an abbreviated version of the fitting analysis.}

\begin{figure*}
    \centering
    \includegraphics[width=\textwidth]{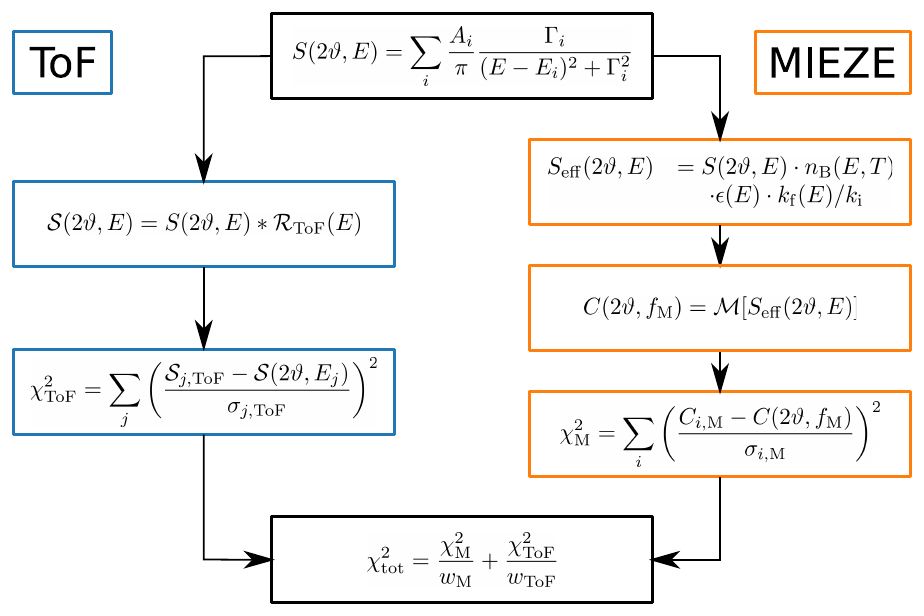}
    \caption{\label{fig:flow_chart} Flow chart of fitting combined data sets. The time-of-flight procedure is depicted using blue boxes, and the MIEZE procedure is highlighted using orange ones. \lukas{indices in der ersten Formel von $i$ zu $n$ ändern}}
\end{figure*}

Starting from this model, the data recorded in pure water at $T = 300\,\mathrm{K}$ using ToF and MIEZE spectroscopy was analyzed via fitting of a combined least-squares function. 
Fig. \ref{fig:flow_chart} summarizes the fitting process in a diagram.
The treatment follows two different pathways for the ToF and the MIEZE data depicted in terms of blue and orange boxes, respectively.

For a proper comparison between the time-of-flight data and the sum of Lorentzian distributions, the resolution function must be included. 
While the elastic linewidth can be measured using a vanadium reference sample, the resolution at finite energy transfers can only be estimated by means of analytical models or simulation. 
In the case of the TOFTOF instrument, this has been explicitly studied by Unruh \textit{et al.} \cite{UNRUH2007} and Gaspar \cite{2007Gaspar}. 
The main contributions to the uncertainty of the energy transfer were found to be the opening angles of the pulsing and monochromatizing choppers, as well as the detector tubes. 
The latter introduced uncertainties in terms of the detector's dead time, differences of flight time due to the geometry, and the wavelength dependent detection efficiency of the tubes. 
The dependence on scattering angle was not included in the resolution function and was neglected in this treatment.

The convolution of the heuristic model with the TOFTOF resolution function is given by
\begin{equation}
    \label{eq:tof_reso_convo}
    \mathcal{S}(2\vartheta, E_j) = \int \frac{1}{\sqrt{2 \pi \sigma_{E_j}}} \exp{\left( \left[ -\frac{E - E_j}{\sqrt{2} \sigma_{E_j}} \right]^2 \right)}\cdot S(2\vartheta, E)\,\mathrm{d}E,
\end{equation}
where $\sigma_{E_j}$ is the standard deviation expressing the instrumental uncertainty evaluated at the energy transfer $E_j$. 
Using this expression, the least-squares function for the TOFTOF data $\chi^2_\mathrm{ToF}$ was computed (see Fig. \ref{fig:flow_chart}). 
For the sake of comparison, both, the data set and convolution models were normalized to unit area such that $\int \mathcal{S}(2\vartheta,E) \mathrm{d}E = 1$.

\begin{figure*}[h]
    \centering
    \includegraphics[width=\textwidth]{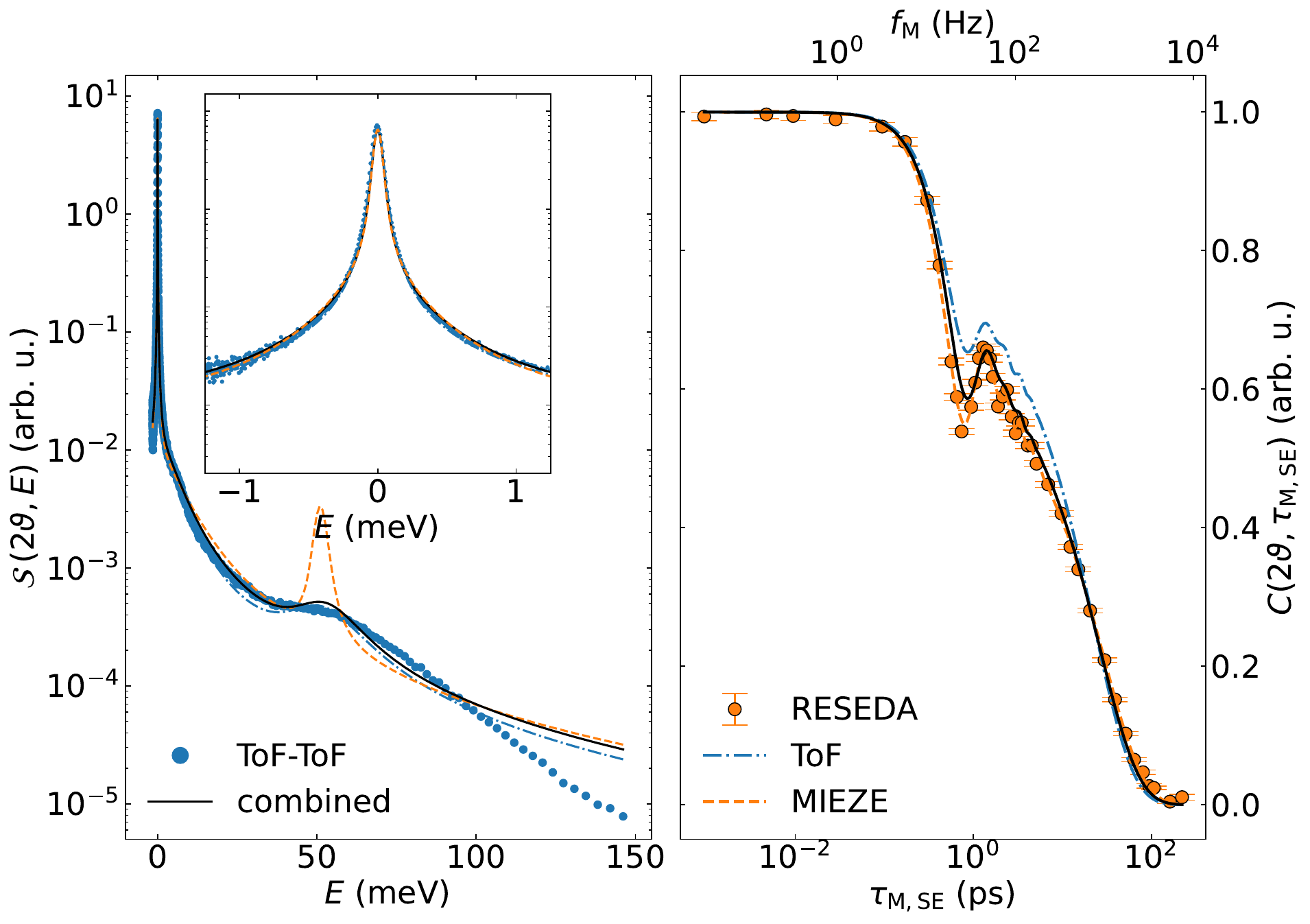}
    \caption{\label{fig:3QENS1INS} Results of fitting the 3Q1I model. The points in blue or red visualize the measurement of TOFTOF and RESEDA, respectively. Wherever errorbars are missing, they would be significantly smaller then the size of the markers. The solid lines represent the result of the combined fit analysis. The dashed orange line represents the fit of the RESEDA data and has been convoluted with the instrument resolution for comparison with the TOFTOF data. Similarly, the blue, dotted line visualizes the fit of the TOFTOF data and has been transformed for comparison with the RESEDA data. The fit of the MIEZE data is insensitive to the inelastic scattering contribution, which is why the amplitude is overestimated \corr{to} improve the at low energy transfers.}
\end{figure*}

The analysis of the MIEZE data required the transformation of the baseline model from the energy domain into the time domain. Especially in the case of large energy transfers, this required computation of the contrast using Eq. \ref{EqCgen}. 
For the baseline model to comply with the theoretical framework laid out in the main text, the detailed balance factor $n_\mathrm{B} (E, T) = \exp{\left( -E / k_\mathrm{B} T \right)}$, the detector efficiency $\epsilon (E)$ and the correction factor $k_\mathrm{f}(E) / k_\mathrm{i}$ were taken into account in the model of the dynamic structure factor. 
This is marked in the upper orange box in Fig. \ref{fig:flow_chart}. 
Subsequently, the transformation, denoted $\mathcal{M}[\cdots]$, was calculated using the effective structure factor ($S_\mathrm{eff}(2\vartheta,E)$) in Eq. \ref{EqCgen}.
Then the least-squares value of the MIEZE data, $\chi^2_\mathrm{M}$ is computed.
\corr{The result of this procedure for the 3Q1I model is shown in Fig.\,\ref{fig:3QENS1INS}}.

Summing up $\chi^2_\mathrm{X}$ for both $\mathrm{X} = \text{ToF, M}$ yields the total least-squares value $\chi^2_\mathrm{tot}$, which should be minimized by the optimal model through variation of the free parameters, $A_i$, $E_i$ and $\Gamma_i$ of each Lorentzian peak.
However, adding up the $\chi^2$ values is not straightforward. 
Due to the comparatively large number of data points and the smaller relative uncertainty, $\chi^2_\mathrm{ToF}$ dominates the combined least-squares function. 
While this has far reaching consequences, especially in the context of the estimated uncertainties, we opted to weight $\chi^2_\mathrm{X}$ by $1/N_\mathrm{X}$,  with $N_\mathrm{X}$ being the number of data points.

The model parameters were optimized numerically using the \textit{differential evolution} algorithm \cite{storn1997differential} implemented in the \textit{scipy} \cite{2020SciPy-NMeth} library as well as the \textit{Python} wrapper package \textit{iminuit}, which makes the \textit{MIGRAD} algorithm of CERN's \textit{ROOT} data analysis framework available \cite{dembinski_iminuit_2020, antcheva_root_2011}. 
The uncertainty of the fit parameters obtained were evaluated using the inverse \textit{Hessian matrix} and population statistics acquired by differential evolution fitting, as in appendix \ref{app:uncertainty}.

\newpage
\section{Uncertainty analysis}
\label{app:uncertainty}

Using one single data set only, one can use the common uncertainty estimation, in terms of a quadratic approximation of the least-squares cost function around the minimum. 
For the least-squares analysis, the $1\sigma$ uncertainty estimate and covariance of a fit parameter $a_i$ is given by the contour in parameter space, where $\chi^2 = \chi^2_\mathrm{min} + 1$. This contour can be traced out by mapping out the least-squares cost function around the minimum, which becomes exceedingly time consuming in cases of a high dimensional fit parameter space. More efficient is the calculation of a quadratic approximation of the least-squares function using the \textit{Hessian matrix} used in \textit{gradient descent} or related minimization algorithms. The standard uncertainty of a fit parameter $a_i$ is then calculated as $\sigma_{a_i} = \left(\mathbf{H}^{-1}\right)_{ii}$. Ideally, the fit of an appropriate model function to a set of data points and their standard deviation $(y_i, \sigma_i)$ should result in a goodness-of-fit value $g_\mathrm{fit} \approx 1$,
which is calculated as
\begin{equation}
\label{eq:goodness-of-fit}
g_\mathrm{fit} = \frac{1}{n - m} \sum^n_i w_i \cdot \left[ f(x_i|\mathbf{a}) - y_i \right]^2~,
\end{equation}
similar to the reduced $\chi^2$ value. However, the $\sigma_i$ values do not necessarily capture all uncertainty associated with the value $y_i$ leading to a $g_\mathrm{fit}$ significantly larger than $1$. $n$ is the number of data points and $m$ is the number of fit parameters included in the model function $f$. In a first attempt to account for this fact, $g_\mathrm{fit}$ can be used as a scaling factor between the weights in the least-squares function $w_i$ and the $\sigma_i$ associated with each data point: $w_i = g_\mathrm{fit} / \sigma_i^2$. Then, $\sigma_i$ represent relative uncertainties between the data points, but not the \textit{'total uncertainty'}. As a result, the uncertainties of the fit parameters need to be re-scaled to $\sigma_{a_i}^2 = g_\mathrm{fit} \cdot \left(\mathbf{H}^{-1}\right)_{ii}$ as well \cite{strutz2011data}.

The method to estimate the reliability of the fit parameters assumes certain statistics of the data points used in the computation of the $\chi^2$ value. In case of the combined model with $\chi^2_\mathrm{tot}$, the unnatural emphasis of the MIEZE data skews the results compared to a standard least-squares function. Thus, besides the uncertainty estimates obtained from the calculations described above, we used the intermediate results of the differential evolution algorithm to study the distribution of fit parameters, which yielded $\chi_\mathrm{tot}^2 \leq \chi^2_\mathrm{tot, min} \cdot 1.02$. The populations of vectors in parameter space, which are drawn by the differential evolution algorithm, have been produced via the \textit{best1bin} strategy \cite{storn1997differential}. The uncertainty of each optimal parameter is then determined by calculating the standard deviation of the marginalized distribution with respect to the optimal value. For a $\SI{2}{\%}$ deviation from the minimal $\chi^2$ between \num{200} and \num{400} parameter vectors are used for this analysis.

\end{document}